\RequirePackage{ifpdf}
\ifpdf % We are running pdfTeX in pdf mode
\documentclass[pdftex]{sigma}
\else
\documentclass{sigma}
\fi

\numberwithin{equation}{section}

\begin{document}

\allowdisplaybreaks

\renewcommand{\PaperNumber}{070}

\FirstPageHeading

\ShortArticleName{$C$-Integrability Test for Discrete Equations via Multiple Scale
Expansions}

\ArticleName{$\boldsymbol{C}$-Integrability Test for Discrete Equations\\ via Multiple Scale
Expansions}

\Author{Christian SCIMITERNA and Decio LEVI}
\AuthorNameForHeading{C.~Scimiterna and D.~Levi}
\Address{Dipartimento di Ingegneria Elettronica, Universit\`a degli Studi Roma Tre and Sezione INFN, Roma Tre, Via della Vasca Navale 84, 00146 Roma, Italy}
\Email{\href{mailto:scimiterna@fis.uniroma3.it}{scimiterna@fis.uniroma3.it}, \href{mailto:levi@roma3.infn.it}{levi@roma3.infn.it}}

\ArticleDates{Received May 29, 2010, in f\/inal form August 20, 2010;  Published online August 31, 2010}

\Abstract{In this paper we are extending the well known  integrability theorems obtained by multiple scale techniques  to the case of linearizable dif\/ference equations. As an example we apply the theory to the case of a dif\/ferential-dif\/ference dispersive  equation of the Burgers hierarchy which via a discrete Hopf--Cole transformation reduces to a linear dif\/ferential dif\/ference equation. In this case  the equation satisf\/ies the $A_1$, $A_2$ and $A_3$ linearizability conditions. We then consider its  discretization. To get a  dispersive equation we  substitute the {\it time} derivative by its symmetric discretization. When we apply to this nonlinear  partial dif\/ference equation the multiple scale expansion we f\/ind out that the lowest order non-secularity condition is given by a non-integrable nonlinear Schr\"odinger equation. Thus showing that this discretized Burgers equation is neither linearizable not integrable.}

\Keywords{linearizable discrete equations; linearizability theorem; multiple scale expansion; obstructions to linearizability; discrete Burgers}

\Classification{34K99; 34E13; 37K10; 37J30}

\section{Introduction}

Calogero in 1991~\cite{ca} introduced the notion of $S$ and $C$ integrable equations to denote those nonlinear partial dif\/ferential equations which are solvable through an inverse {\bf S}pectral transform or linearizable through a {\bf C}hange of variables. Using the multiple scale reductive technique he was able to show that the Nonlinear Schr\"odinger Equation (NLSE)
\begin{gather}
 \mbox{i} \partial_t f + \partial_{xx} f = \rho_{2} |f|^2 f,\qquad f=f(x,t), \label{e0}
\end{gather}
appears as a universal equation governing the evolution of slowly varying packets of quasi-monochromatic waves in weakly nonlinear media featuring dispersion. Calogero and Eckhaus then showed that  a necessary condition for the $S$-integrability of a dispersive nonlinear partial dif\/ferential equation is that its multiple scale expansion around a slowly varying packet of quasi-monochromatic wave  should provide at the lowest order in the perturbation parameter an integrable NLSE. Then, by a paradox, they showed that a $C$-integrable equation  must reduce to a linear equation or another $C$-integrable equation as the Eckhaus equation \cite{ce,ls2009}.

In the case of discrete equations it has been shown \cite{Schoo,Ag,lm,LevHer,levi,lp,HLPS,HLPS2,HLPS3} that a similar situation is also true. One presents the equivalent of the Calogero--Eckhaus theorem stating that {a necessary condition for} a nonlinear dispersive partial dif\/ference equation to be $S$-integrable is that the lowest order multiple scale expansion on $\mathcal{C}^{(\infty)}$ functions  give rise to integrable NLSE. The nonlinear dispersive partial dif\/ference equation will be $C$-integrable if its multiple scale expansion on $\mathcal{C}^{(\infty)}$ functions will give rise at the lowest order to a linear or a~$C$-integrable dif\/ferential equation.

By going to a higher order in the expansion, the multiple scale techniques {give more stringent conditions which} have been used  to f\/ind new $S$-integrable Partial Dif\/ferential Equations (PDE's) and   to prove the  integrability of new nonlinear equations~\cite{DMS,DP,MK}. Probably the most important example of such nonlinear PDE is  the Degasperis--Procesi equation~\cite{dp1}. Up to our knowledge  higher order expansions for nonlinear  linearizable equations have not been considered in details.

The purpose of this paper is to show that the integrability theorem, stated in~\cite{HLPS2}, can be extended to the case of linearizable dif\/ference equations, providing a way to discriminate between $S$-integrable, $C$-integrable and non--integrable lattice equations. The continuous (and thus discrete) higher order $C$-integrability conditions are, up to our knowledge, presented here for the f\/irst time. We apply here the resulting linearizability conditions to a dif\/ferential-dif\/ference dispersive nonlinear equation of the discrete Burgers hierarchy~\cite{BLR} and  its dif\/ference-dif\/ference analogue.

In Section~\ref{section2} we  present the dif\/ferential-dif\/ference linearizable nonlinear dispersive Burgers and its partial dif\/ference analogue and discuss the tools necessary to carry out the multiple scale $C$-integrability test. In Section~\ref{section3}  we apply them to the two  equations previously introduced, leaving to the Appendix all details of the calculation, and present in Section~\ref{section4} some conclusive remarks.

\section{Multiple scale perturbation reduction of Burgers equations}\label{section2}

The Burgers equation, the simplest nonlinear equation for the study of gas dynamics with heat conduction and viscous ef\/fect, was introduced by Burgers in 1948~\cite{burgers,wh}.
Explicit solutions of the Cauchy problem on the inf\/inite line for the Burgers equation may be obtained by  the
Hopf--Cole transform, introduced independently by Hopf and Cole in 1950~\cite{cole,hopf}.
This transformation  linearize the equation and the solution of the linearized equation provide solutions of the Burgers equation.

Bruschi, Levi and Ragnisco \cite{BLR} extended the Hopf--Cole transformation to  construct hierarchies of linearizable nonlinear matrix PDE's, nonlinear dif\/ferential-dif\/ference equations and dif\/ference-dif\/ference equations.

The simplest dif\/ferential--dif\/ference nonlinear dispersive equation of the Burgers hierarchy is given by
\begin{gather}
\partial_{t}u_{n}\left(t\right)=\frac{1} {2h}\left\{\left[1+hu_{n}\left(t\right)\right]\left[u_{n+1}\left(t\right)-u_{n}\left(t\right)\right]-\frac{u_{n-1}\left(t\right)-u_{n}\left(t\right)} {1+hu_{n-1}\left(t\right)}\right\},\label{Numa}
\end{gather}
where the function $u_{n}(t)$ and the lattice parameter  $h$ are all supposed to be real.
Equation~(\ref{Numa}) has a nonlinear dispersion relation $\omega=-\frac{\sin\left(\kappa h\right)}{h}$.

When $h\rightarrow 0$ and $n \rightarrow \infty$ in such a way that $x=nh$ is f\/inite,  the Burgers equation (\ref{Numa}) reduces to the one dimensional wave equation $\partial_{t}u-\partial_{x}u=\mathcal O(h^3)$.

Through the discrete Cole--Hopf transformation
\begin{gather}
u_{n}\left(t\right)=\frac{\phi_{n+1}\left(t\right)-\phi_{n}\left(t\right)} {h\phi_{n}\left(t\right)} \label{Reghini}
\end{gather}
equation (\ref{Numa}) linearizes to the discrete linear wave equation
\begin{gather} \label{a2}
\partial_{t}\phi_{n}\left(t\right)=\frac{\phi_{n+1}\left(t\right)-\phi_{n-1}\left(t\right)} {2h}.
\end{gather}
The transformation (\ref{Reghini}) can be inverted and gives
\begin{subequations}\label{d}
\begin{gather}
\phi_{n}=\phi_{n_0}\prod_{j=n_0}^{j=n-1}\left(1+hu_{j}\right),\qquad n\geq n_0+1,\\
\phi_{n}=\frac{\phi_{n_0}}{\prod\limits_{j=n}^{j=n_0-1}\left(1+hu_{j}\right)},\qquad n\leq n_0-1,
\end{gather}
\end{subequations}
where $\phi_{n_0}=\phi_{n_0}(t)$ is  the function $\phi_{n}$ calculated at a given initial point $n=n_0$.  When $u_{n}(t)$ satisf\/ies  equation~(\ref{Numa}), the function $\phi_{n}(t)$ will satisfy the discrete wave equation (\ref{a2})  if $\phi_{n_0}$  satisf\/ies the ordinary dif\/ferential equation
\[
\dot\phi_{n}-\frac{1} {2h}\left[1+hu_{n}-\frac{1} {\left(1+hu_{n-1}\right)}\right]\phi_{n}\Bigl |_{n=n_0}=0,
\]
whose solution is given by
\[
\phi_{n_0}(t)=\phi_{n_0}(t_0)\exp\left\{\frac{1} {2h}\int_{t_0}^{t}\left[1+hu_{n}-\frac{1} {\left(1+hu_{n-1}\right)}\right]\Bigl |_{n=n_0}dt^{\prime}\right\},
\]
 and $t_0$ is an initial  time.
 If $\lim\limits_{n\to -\infty} u_{n}(t)=u_{-\infty}$ is f\/inite equations (\ref{d}) reduce to
\begin{gather*}
\phi_{n}=\alpha\left(t\right)\left(1+hu_{-\infty}\right)^n\prod_{\gamma=-\infty}^{\gamma=n-1}\left(\frac{1+hu_{\gamma}} {1+hu_{-\infty}}\right),
\end{gather*}
where $\alpha\left(t\right)$ is a $t$-dependent function,
\begin{gather}\nonumber
\alpha(t)=\alpha_{0}\exp\left\{\frac{1} {2}\left(1+\frac{1} {1+hu_{-\infty}}\right)u_{-\infty}t\right\}.
\end{gather}

A $C$-integrable discretization of  equation (\ref{Numa}) is given by the partial dif\/ference equation
\begin{gather}\label{bd}
\frac{u_{n,m+1}-u_{n,m}}{\sigma}=\frac{1}{2h}\left[ (1+h u_{n,m})(u_{n+1,m}-u_{n,m+1}) -\frac{u_{n-1,m}-u_{n,m+1}}{1+hu_{n-1,m}}\right],
\end{gather}
where $\sigma$ is the constant  lattice parameter in the time variable. Equation (\ref{bd}) {is dissipative as it} has a complex dispersion relation $\omega=\frac{i}{\sigma}\ln\left[1 + i \frac{\sigma}{h}\sin\left(\kappa h\right)\right]$.
 As we are {\it not} able to construct a~dispersive counterpart of equation~(\ref{Numa}) we  consider a straightforward discretization of equation~(\ref{Numa})
\begin{gather}
\frac{u_{n,m+1}-u_{n,m-1}} {2\sigma}=\frac{1} {2h}\left[\left(1+hu_{n,m}\right)\left(u_{n+1,m}-u_{n,m}\right)-\frac{\left(u_{n-1,m}-u_{n,m}\right)} {1+hu_{n-1,m}}\right],\label{Pitagora}
\end{gather}
{whose nonlinear} dispersion relation   is
\[
\sin\left(\omega\sigma\right)=-\frac{\sin\left(\kappa h\right)\sigma} {h}.
\]

In the {remaining part of this section} we will present the tools necessary to carry out the multiple scale expansion of equations (\ref{Numa}), (\ref{Pitagora}) and construct the conditions which they must satisfy to be $C$-integrable equation of $j$ order, with $j=1,2,3$, i.e.\ such that asymptotically { they reduce to a linear equation up to terms respectively of the third, fourth and f\/ifth order in the perturbation paramether.} These conditions up to our knowledge have been presented for the f\/irst time by Dr.\ Scimiterna in his PhD Thesis~\cite{tS} and  are published here for the f\/irst time.

\subsection{Expansion of real dispersive partial dif\/ference equations}

For completeness we brief\/ly illustrate here all the ingredients of the  reductive perturbative technique necessary to treat dif\/ference equations, as presented in~\cite{HLPS, HLPS2,tS}.

\subsubsection{From shifts to derivatives}

Let us consider a function $u_{n}: \mathbb{Z}\rightarrow\mathbb{R}$ depending on an index $n\in\mathbb{Z}$ and let us suppose that:
\begin{itemize}\itemsep=0pt
\item The dependence of $u_{n}$ on $n$ is realized through the \emph{slow variable} $n_{1}\doteq\epsilon n\in\mathbb{R}$, $\epsilon\in\mathbb{R}$, $0<\epsilon\ll 1$, that is to say $u_{n}\doteq u(n_{1})$.
\item The function $u\left(n_{1}\right)\in\mathcal{C}^{(\infty)}\left(\mathcal{D}\right)$, where $\mathcal{D}\in\mathbb{R}$ is a  region containing the point $ n_{1}$.
\end{itemize}
Under these hypotheses one can write the action of the shift operator $T_{n}$ such that $T_{n} u_{n}\doteq u_{n+1}=u(n_{1}+\epsilon)$  as the following (formal) series
\begin{gather}
T_{n} u(n_{1})=u( n_{1})+\epsilon u_{,n_1}( n_{1})+\frac{\epsilon^2}{2
}u_{,2\,n_1}( n_{1})+ \cdots +\frac{\epsilon^k}{k!}u_{,k\,n_1}( n_{1})+ \cdots\nonumber\\
\hphantom{T_{n} u(n_{1})}{}
=\sum_{k=0}^{+\infty}\frac{\epsilon^k}{k!} u_{,k\,n_1}( n_{1}),\label{Larth}
\end{gather}
where $u_{,k\,n_1}( n_{1})\doteq d^{k} u(n_{1}) /dn_{1}^{k}\doteq d_{n_{1}}^{k} u( n_{1})$, being $d_{n_{1}}$ the total derivative operator. The last expression suggests the following formal expansion for the dif\/ferential operator $T_n$:
\begin{gather}
T_{n}=\sum_{k=0}^{+\infty}\frac{\epsilon^k}{k!}\,d_{n_{1}}^k\doteq e^{\epsilon d_{n_{1}}} \nonumber
\end{gather}
valid only when the series in equation~(\ref{Larth}) is converging.  So we  must require that the radius of convergence of the series starting at $n_{1}$ is wide enough to include as an \emph{inner point} at least the point $ n_{1}+\epsilon$.

Let us introduce more complicated dependencies of $u_{n}$ on $n$. For example one can assume a~simultaneous dependence on the \emph{fast variable} $n$ and on the slow variable $n_{1}$, i.e.\ $u_{n}\doteq u(n,n_{1})$. The action of the \emph{total} shift operator $T_{n}$ will now be given by $T_{n} u_{n}\doteq u_{n+1}=u(n+1,n_{1}+\epsilon)$ so that we can write $T_{n}\doteq\mathcal{T}_{n}^{(1)}\mathcal{T}_{n_{1}}^{(\epsilon)},
$ where the \emph{partial} shift operators $\mathcal{T}_{n}^{(1)}$ and $\mathcal{T}_{n_{1}}^{(\epsilon)}$ are def\/ined respectively by
\[
\mathcal{T}_{n}^{(1)}u(n,n_{1})=u(n+1,n_{1})=\sum_{k=0}^{\infty} \frac{1}{k!}\partial_{n}^k u(n,n_1) = e^{ \partial_{n}}u(n,n_1),
\]
and
\begin{gather}\label{11b}
\mathcal{T}_{n_{1}}^{(\epsilon)}u(n,n_{1})=u(n,n_{1}+\epsilon)=\sum_{k=0}^{\infty} \frac{\epsilon^k}{k!}\partial_{n_1}^k u(n,n_1) = e^{\epsilon \partial_{n_1}}u(n,n_1).
\end{gather}
 The dependence of $u_{n}$ on $n$ can be easily extended to the case of one fast variable $n$ and $K$ slow variables $n_{j}\doteq\epsilon_{j}n$, $\epsilon_{j}\in\mathbb{R}$, $1\leq j\leq K$ each of them being def\/ined by its own parameter $\epsilon_{j}$. The action of the total shift operator $T_n$ will now be given in terms of the partial shifts $\mathcal{T}_{n}^{(1)}$, $\mathcal{T}_{n_{j}}^{(\epsilon)}$, as
$T_{n}\doteq\mathcal{T}_{n}^{(1)}\prod\limits_{j=1}^K\mathcal{T}_{n_{j}}^{(\epsilon_{j})}$.

Let us now consider a nonlinear partial dif\/ference equation
\begin{gather}
F\left[\left \{u_{n+k,m+j}
\right\}^{j=\left (-{\cal K}^{(-)},  \mathcal{K}^{(+)}\right)}_{k=\left (-\mathcal{N}^{(-)},  {\cal N}^{(+)}\right)}\right]=0,\qquad \big(\mathcal{N}^{(\pm)},  \mathcal{K}^{(\pm)}\big)\geq 0,\label{Esperia}
\end{gather}
 for a function $u_{n,m}: \mathbb{Z}^2\rightarrow\mathbb{R}$ which now depends on two indexes $n$ and $m\in\mathbb{Z}$ which we will term respectively as discrete \emph{space} and \emph{time} indices. Equation~(\ref{Esperia}) contains $m$ and $n$-shifts, respectively in the intervals $(m-{\mathcal{K}}^{(-)},   m+{\mathcal{K}}^{(+)})$ and $(n-{\mathcal{N}}^{(-)},  n+{\mathcal{N}}^{(+)})$. Under some obvious hypothesis on the $\mathcal{C}^{(\infty)}$ property of the function $u_{n,m}$ and on the radius of convergence of its Taylor expansion for all  shifts in the indices $n$ and $m$ involved in the dif\/ference equation (\ref{Esperia}), we can write a series representation of  $u_{n+k,m+j}$ around  $u_{n,m}$. We choose the slow variables as $n_{k}\doteq\epsilon_{n_{k}}n$, $m_{j}\doteq\epsilon_{m_{j}}m$ with
\begin{gather}
\nonumber\epsilon_{n_{k}}\doteq N_{k}\epsilon^{k},\qquad 1\leq k\leq K_{n},\qquad \epsilon_{m_{j}}\doteq M_{j}\epsilon^{j},\qquad 1\leq j\leq K_{m},
\end{gather}
where the various constants $N_{k}$, $M_{j}$ and $\epsilon$ are all real numbers. In this presentation we assume $K_{n}=1$ and $K_{m}=K$ (eventually $K=+\infty$) so that
\begin{subequations}\label{Istria}
\begin{gather}
T_{n}=\mathcal{T}_{n}^{(1)}\mathcal{T}_{n_{1}}^{(\epsilon_{n_{1}})}
=\mathcal{T}_{n}^{(1)}\sum_{j=0}^{+\infty}\epsilon^j{\mathcal{A}}_{n}^{(j)},\label{Istria1}\\
T_{m}=\mathcal{T}_{m}^{(1)}\prod_{j=1}^K\mathcal{T}_{m_{j}}^{(\epsilon_{m_{j}})}
=\mathcal{T}_{m}^{(1)}\sum_{j=0}^{+\infty}\epsilon^j{\mathcal{B}}_{m}^{(j)},\\
T_{n}T_{m}=\mathcal{T}_{n}^{(1)}\mathcal{T}_{m}^{(1)}\mathcal{T}_{n_{1}}^{(\epsilon_{n_{1}})}
\prod_{j=1}^K\mathcal{T}_{m_{j}}^{(\epsilon_{m_{j}})}
=\mathcal{T}_n^{(1)}\mathcal{T}_m^{(1)}\sum_{j=0}^{+\infty}\epsilon^j{\mathcal{C}}_{n,m}^{(j)},
\end{gather}
\end{subequations}
where the operators   ${\mathcal{A}}_{n}^{(j)}$, ${\mathcal{B}}_{m}^{(j)}$, and ${\mathcal{C}}_{n,m}^{(j)}$ are given in  Table~\ref{table1}.
Inserting the explicit expressions~(\ref{Istria}) of the shift operators into equation~(\ref{Esperia}), this turns into a PDE of inf\/inite order.
So we will assume for the function $u_{n,m}=u(n,m,n_{1},\left\{m_{j}\right\}_{j=1}^{K},\epsilon)$ a double expansion in harmonics and in the perturbative parameter $\epsilon$
\begin{gather}
u_{n,m}=\sum_{\gamma=1}^{+\infty}\sum_{\theta=-\gamma}^{\gamma}\epsilon^{\gamma}u_{\gamma}^{(\theta)}\left(n_{1},m_{j},j\geq 1\right)e^{i\theta\left(\kappa hn-\omega(\kappa) \sigma m\right)},\label{Catilina}
\end{gather}
with $u_{\gamma}^{(-\theta)}\left(n_{1},m_{j},j\geq 1\right)=\bar u_{\gamma}^{(\theta)}\left(n_{1},m_{j},j\geq 1\right)$, where by a bar we denote the complex conjugate,  in order to ensure the reality of $u_{n,m}$. The index $\gamma$ is chosen $\geq 1$ so that the nonlinear terms of equation~(\ref{Esperia}) enter as a perturbation in the multiple scale expansion. For simplicity we will set $N_{1}=M_{j}=1$, $j\geq 1$. Moreover we will assume that the functions $u_{\gamma}^{(\theta)}$ satisfy the asymptotic conditions $\lim\limits_{n_{1}\rightarrow\pm\infty}u_{\gamma}^{(\theta)}=0$, $\forall\,\gamma$ and $\theta$ to provide a meaningful expansion.

\begin{table}[t]
\centering\footnotesize
\caption{The operators ${\mathcal{A}}_{n}^{(j)}$, ${\mathcal{B}}_{m}^{(j)}$ and ${\mathcal{C}}_{n,m}^{(j)}$ appearing in equations (\ref{Istria}).}\label{table1}
\vspace{1mm}

\begin{tabular}{||@{\,}c@{\,}||@{\,}c@{\,}|@{\,}c@{\,}|@{\,}c@{\,}|@{\,}c@{\,}|@{\,}c@{\,}||}
 \hline
 & $j=0$ & $j=1$ & $j=2$ & $j=3$ & $j=4$\\
 \hline\hline
\tsep{1.5ex}\bsep{0.5ex}
${\mathcal{A}}_{n}^{(j)}$ & $1$ & $N_{1}\partial_{n_{1}}$ & $\frac{N_{1}^2} {2}\partial_{n_{1}}^2$ & $\frac{N_{1}^3} {6}\partial_{n_{1}}^3$ & $\frac{N_{1}^4} {24}\partial_{n_{1}}^4 $ \\\hline
\tsep{1.5ex}
${\mathcal{B}}_{m}^{(j)}$ & $1$ & $M_{1}\partial_{m_{1}}$ & $\frac{M_{1}^2} {2}\partial_{m_{1}}^2+M_{2}\partial_{m_{2}}$ & $\frac{M_{1}^3} {6}\partial_{m_{1}}^3+$  & $\frac{M_{1}^4} {24}\partial_{m_{1}}^4+\frac{M_{1}^2M_{2}} {2}\partial_{m_{1}}^2\partial_{m_{2}}+$\\
 & & & & $ +M_{1}M_{2}\partial_{m_{1}}\partial_{m_{2}}+$ &  $+\frac{M_{2}^2} {2}\partial_{m_{2}}^2+ M_{1}M_{3}\partial_{m_{1}}\partial_{m_{3}}+M_{4}\partial_{m_{4}}$\\
 &&& &$+ M_{3}\partial_{m_{3}} $&
\bsep{0.5ex} \\\hline
\tsep{1.5ex}
${\mathcal{C}}_{n,m}^{(j)}$ & $1$ & ${\mathcal{A}}_{n}^{(1)}+{\mathcal{B}}_{m}^{(1)}$ & ${\mathcal{A}}_{n}^{(2)}+{\mathcal{B}}_{m}^{(2)}+$ & ${\mathcal{A}}_{n}^{(3)}+{\mathcal{B}}_{m}^{(3)}+$ & ${\mathcal{A}}_{n}^{(4)}+{\mathcal{B}}_{m}^{(4)}+$\\
 & & & $+N_{1}M_{1}\partial_{n_{1}}\partial_{m_{1}}$ & $+N_{1}M_{2}\partial_{n_{1}}\partial_{m_{2}}+$ & $+\frac{M_{1}^3N_{1}} {6}\partial_{m_{1}}^3\partial_{n_{1}}+\frac{N_{1}^3M_{1}} {6}\partial_{n_{1}}^3\partial_{m_{1}}+$\\
 & & & &
 $+\frac{M_{1}N_{1}^2} {2}\partial_{n_{1}}^2\partial_{m_{1}}+ $
 & $+N_{1}M_{1}M_{2}\partial_{n_{1}}\partial_{m_{1}}\partial_{m_{2}}+\frac{N_{1}^2M_{2}} {2}\partial_{n_{1}}^2\partial_{m_{2}}+$\\
 & & & & $+\frac{N_{1}M_{1}^2} {2}\partial_{n_{1}}\partial_{m_{1}}^2$ & $+\frac{N_{1}^2M_{1}^2} {4}\partial_{n_{1}}^2\partial_{m_{1}}^2
 +N_{1}M_{3}\partial_{n_{1}}\partial_{m_{3}}$ \bsep{0.5ex} \\ \hline
\end{tabular}
\end{table}

\subsubsection{From derivatives to shifts}

The multiple scale approach discussed above reduces a given partial dif\/ference equation into a~partial dif\/ferential equation for  the amplitudes $u^{(\theta)}_{\gamma}$ contained in the def\/inition (\ref{Catilina}).

We can rewrite the so obtained partial dif\/ferential equation  as a partial dif\/ference equation inverting the expansion of the partial shift operator in term of partial derivatives (\ref{11b}). From~(\ref{11b}) we have
\begin{gather}\label{i1}
\partial_{n_{1}}=\frac{1}{\epsilon}\ln\mathcal{T}_{n_{1}}^{(\epsilon)}=
\frac{1}{\epsilon}\ln\big(1+\epsilon\Delta^{(+)}_{n_{1}}\big)\doteq\sum_{k=1}^{+\infty}\frac{(-\epsilon)^{k-1}}{k} \big[\Delta^{(+)}_{n_{1}}\big]^k,
\end{gather}
where $\Delta^{(+)}_{n_{1}}\doteq(\mathcal{T}_{n_{1}}^{(\epsilon)}-1)/\epsilon$ is the f\/irst \emph{forward} dif\/ference operator with respect to the slow-variable~$n_{1}$. This is just one of the possible inversion formulae for the operator $\mathcal{T}_{n_{1}}^{(\epsilon)}$. For example an  expression similar to equation~(\ref{i1}) can be written for the f\/irst \emph{backward} dif\/ference operator $\Delta^{(-)}_{n_{1}}\doteq \big(1-\big[\mathcal{T}_{n_{1}}^{(\epsilon)} \big]^{-1}\big)/\epsilon$. For the f\/irst \emph{symmetric} dif\/ference operator $\Delta^{(s)}_{n_{1}}\doteq\big( \mathcal{T}_{n_{1}}^{(\epsilon)}-\big[\mathcal{T}_{n_{1}}^{(\epsilon)} \big]^{-1}\big)/2\epsilon$ we get{\samepage
\[
\partial_{n_{1}}=\sinh^{-1}\epsilon\Delta^{(s)}_{n_{1}}\doteq\sum_{k=1}^{+\infty}\frac{P_{k-1}(0) \epsilon^k} {k}\big[\Delta^{(s)}_{n_{1}}\big]^k,
\]
where $P_{k}(0)$ is the $k$-th \emph{Legendre} polynomial evaluated at $x=0$.}

Only when we impose that the function $u_n$ is a slow-varying function of order $\ell$ in the variable $n_{1}$, i.e.\ that $\Delta^{\ell+1}_{n_{1}}u_n=0$, we can see that the $\partial_{n_{1}}$ operator, which is given by formal series containing in general inf\/inite powers of the $\Delta_{n_{1}}$, reduces to polynomial of order at most~$\ell$. In~\cite{levi}, choosing $\ell=2$ for the indexes $n_{1}$ and $m_{1}$ and $\ell=1$ for $m_{2}$, it was shown that the integrable \emph{lattice potential KdV} equation~\cite{31} reduces to a completely discrete and local nonlinear Schr\"odinger equation which has been proved to be not integrable by singularity conf\/inement and algebraic entropy \cite{viallet,ramani}. Consequently, if one passes from derivatives to shifts, one ends up in general with a \emph{nonlocal} partial dif\/ference equation in the slow variables $n_{\kappa}$ and $m_{\delta}$.

\subsection[The orders beyond the Schr\"odinger equation and the $C$-integrability conditions]{The orders beyond the Schr\"odinger equation\\ and the $\boldsymbol{C}$-integrability conditions}

The multiple scale expansion of an  equation  of the Burgers hierarchy on functions of inf\/inite order will thus give rise to  PDE's. So a multiple scale integrability test will require that a~dispersive equation like equation~(\ref{Numa})  is $C$-integrable if its multiple scale expansion will go into the hierarchy of the Schr\"odinger equation
\[
i\partial_{t}\psi+\partial_{x}^{2}\psi=0.
\]
 To be able to verify the $C$-integrability  we need to consider  in principle all the orders beyond  the Schr\"odinger equation. This in general will not be possible but already a few orders beyond  the Schr\"odinger equation might be suf\/f\/icient to verify if the equation is linearizable or not.  In the case of $S$-integrable nonlinear PDE's the f\/irst attempt to go beyond the NLSE order has been presented by  Degasperis, Manakov and Santini in \cite{DMS}. These authors, starting from an $S$-integrable model, through a combination of an asymptotic functional analysis and spectral methods, succeeded in removing all the secular terms from the reduced equations order by order. Their results could be summarized as follows:

\begin{enumerate}\itemsep=0pt

\item The number of slow-time variables required for the amplitudes $u^{\left(\theta\right)}_{j}$  appearing in (\ref{Catilina}) coincides with the number of nonvanishing coef\/f\/icients of the Taylor expansion of the dispersion relation,  $\omega_{j}\left(\kappa\right)=\frac{1}{j!}\frac{d^j \omega(k)}{dk^j}$.

\item The amplitude $u^{\left(1\right)}_{1}$ evolves at the slow-times $m_{s}$, $s\geq 2$ according to the $s$-th equation of the NLS hierarchy.

\item The amplitudes of the higher perturbations of the f\/irst harmonic $u^{\left(1\right)}_{j}$, $j\geq 2$ evolve, taking into account some \emph{asymptotic boundary conditions}, at the slow-times $m_{s}$, $s\geq 2$ according to certain \emph{linear, nonhomogeneous} equations.
\end{enumerate}
Then  they  concluded  that the cancellation at each stage of the perturbation process of all the secular terms  is a suf\/f\/icient condition  to uniquely f\/ix the evolution equations followed by every~$u^{\left(1\right)}_{j}$, $j\geq 1$ for each slow-time $m_s$. Point~2 implies that a hierarchy of integrable equations provide for a function~$u$ always compatible evolutions, i.e.\ the equations in its hierarchy are \emph{generalized symmetries} of each other.
In this way this procedure provides \emph{necessary and sufficient} conditions to get secularity-free reduced equations \cite{DMS}.

We apply the present procedure to the case of $C$-integrable partial dif\/ference equations.
Following Degasperis and Procesi \cite{DP} we state the following theorem:

\begin{theorem}\label{theorem1}
 If equation \eqref{Esperia} is $C$-integrable then, after a multiple scale expansion, the functions $u^{\left(1\right)}_{j}$, $j\geq1$ satisfy the equations
\begin{subequations}\label{Valentia}
\begin{gather}
\partial_{m_{s}}u^{\left(1\right)}_{1}-(-i)^{s-1}B_{s}\partial_{n_{1}}^{s}u^{\left(1\right)}_{1}\doteq M_{s}u_1^{(1)}=0,\label{Valentia1}\\
M_{s}u^{\left(1\right)}_{j}=f_{s}(j),\label{Valentia2}
\end{gather}
\end{subequations}
$\forall\, j,\, s\geq 2$, where $B_{s}\partial_{n_{1}}^{s}u^{\left(j\right)}_{1}$ is the $s$-th f\/low in the linear Schr\"odinger hierarchy and $B_{s}$ are real constants. All the other  $u_{j}^{(\kappa)}$, $\kappa\geq 2$ are expressed as differential monomials of $u_{r}^{(1)}$, $r \le j-1$.
\end{theorem}

In equation (\ref{Valentia2})  $f_{s}(j)$ is a nonhomogeneous \emph{nonlinear} forcing  written in term of dif\/ferential monomials of $u_{r}^{(1)}$, $r\leq j$. From Theorem~\ref{theorem1} it follows  that  a nonlinear partial dif\/ference equation is said to be $C$-integrable if its asymptotic multiple scale expansion is given by a uniform asymptotic series whose leading harmonic
$u^{(1)}$ possesses  an inf\/inity of generalized symmetries evolving at dif\/ferent times and given by commuting linear equations.
Equations~(\ref{Valentia}) are a~\emph{necessary} condition for $C$-integrability.

It is worthwhile to stress here the non completely obvious fact that, in contrast to  the f\/irst  order  wave equation, $\partial_t u = \partial_x u$, all the symmetries of the Schr\"odinger equation  commuting with it are given by the  equations~(\ref{Valentia1}) and only by them. This implies that all the equations appearing in the multiple scale expansion for a $C$-integrable equation are uniquely def\/ined.

It is obvious that the operators $M_{s}$ def\/ined in equation~(\ref{Valentia1}) commute among themselves.
However the compatibility of equations (\ref{Valentia2}) is not always guaranteed but is subject to some compatibility conditions among their r.h.s.\ terms $f_{s}(j)$.  Once we f\/ix the index $j\geq 2$ in the set of equations~(\ref{Valentia2}), this commutativity condition implies the \emph{compatibility} conditions
\begin{gather}
M_{s}f_{s'}\left(j\right)=M_{s'}f_{s}\left(j\right),\qquad \forall\, s,s'\geq 2,\label{Lavinia}
\end{gather}
where, as $f_{s}\left(j\right)$ and $f_{s'}\left(j\right)$ are functions of the dif\/ferent perturbations $u_r^{(1)}$ of the fundamental harmonic up to degree $j-1$, the time derivatives $\partial_{m_{s}}$, $\partial_{m_{s'}}$  appearing respectively in~$M_{s}$ and~$M_{s'}$ have to be eliminated using the evolution equations (\ref{Valentia}) up to the index $j-1$. These last commutativity conditions turn out to be a~{\it linearizability test}.

Following \cite{DMS} we conjecture that the relations (\ref{Valentia}) are a \emph{sufficient} condition for the $C$-integrability or that the $C$-integrability is a \emph{necessary} condition to have a multiple scale expansion where equations~(\ref{Valentia}) are satisf\/ied. To characterize the functions $f_s(j)$ we introduce the following def\/initions:

\begin{definition}
A dif\/ferential monomial ${\mathcal{M}}\big[u^{\left(1\right)}_{j}\big]$, $j\geq 1$ in the functions $u^{\left(1\right)}_{j}$, its complex conjugate and its $n_1$-derivatives is a monomial of ``gauge'' 1 if it possesses the transformation property
\[
{\mathcal{M}}\big[\tilde u^{\left(1\right)}_{j}\big]= e^{i\theta}{\mathcal{M}}\big[u^{\left(1\right)}_{j}\big], \qquad \mbox{when} \quad \tilde u^{\left(1\right)}_{j}\doteq e^{i\theta}u^{\left(1\right)}_{j}.
\]
\end{definition}

\begin{definition}\label{FAVL1}
A f\/inite dimensional vector space $\mathcal P_{\nu}$, $\nu\geq 2$ is the set of all dif\/ferential polynomials of gauge 1 in the functions $u^{\left(1\right)}_{j}$, $j\geq 1$, their complex conjugates and their $n_1$-derivatives such that their total  order  in $\epsilon$  is  $\nu$, i.e.
\[
\mbox{order}\big(\partial_{n_1}^{\mu}u^{\left(1\right)}_{j}\big)=\mbox{order}\big(\partial_{n_1}^{\mu}\bar u^{\left(1\right)}_{j}\big)=\mu+j=\nu,\qquad \mu\geq 0.
\]
\end{definition}

\begin{definition}\label{FAVL2}
$\mathcal P_{\nu}(\mu)$, $\mu\geq 1$ and $\nu\geq 2$ is the subspace of $\mathcal P_{\nu}$ whose elements are dif\/ferential polynomials of gauge 1  in the functions $u^{\left(1\right)}_{j}$, their complex conjugates and their $n_1$-derivatives such that their total order is $\nu$ and  $1\leq j\leq \mu$.
\end{definition}

From Def\/inition~\ref{FAVL2} it follows that $\mathcal P_{\nu}=\mathcal P_{\nu}(\nu{-}2)$. Moreover  in general $f_{s}(j)\in\mathcal P_{j+s}(j{-}1)$ where $j,\, s\geq 2$. The basis monomials of the spaces $\mathcal P_{\nu}(\mu)$ in which we can express the functions~$f_s(j)$ can be found, for example, in \cite{tS}.

\begin{proposition}
If for each fixed $j\geq 2$ the equation \eqref{Lavinia} with $s=2$ and $s'=3$, namely
\begin{gather}
M_{2}f_{3}\left(j\right)=M_{3}f_{2}\left(j\right), \label{Turno}
\end{gather}
is satisfied, then there exist unique differential polynomials $f_{s}(j)$ $\forall\, s\geq 4$ such that the flows $M_{s}u^{\left(1\right)}_{j}=f_{s}\left(j\right)$ commute for any $s\geq 2$~{\rm \cite{santini2010,degasperisprivate}}.
\end{proposition}

Hence among the relations (\ref{Lavinia}) only those with $s=2$ and $s'=3$ have to be tested.
\begin{proposition}
 The homogeneous equation $M_{s}u=0$ has no solution $u$ in the vector space $\mathcal P_{\mu}$, i.e.\ ${\rm Ker} \left(M_{s}\right)\cap\mathcal P_{\mu}=\varnothing$.
\end{proposition}

 Consequently the multiple scale expansion (\ref{Valentia}) is \emph{secularity-free}. This does not mean that, in solving equation~(\ref{Valentia2}), we have to set to zero all the contributions to the solution coming from the homogeneous equation but only that part of it which is present in the forcing terms. Finally:

\begin{definition}\label{FrancescoColonnaRomano}
If the relations \eqref{Lavinia} are satisf\/ied up to the index $j$, $j\geq 2$, we say that our equation is asymptotically $C$-integrable of degree $j$ or $A_{j}$ $C$-integrable.
\end{definition}

\subsubsection{Integrability conditions for the Schr\"odinger hierarchy}

\indent Here we present the conditions for the asymptotic $C$-integrability of order $k$ or $A_{k}$ $C$-integrability conditions with $k=1,2,3$. To simplify the notation, we will use for $u^{\left(1\right)}_{j}$ the concise form $u(j)$.

The $A_{1}$ $C$-integrability condition is given by the absence of the coef\/f\/icient $\rho_{2}$ of the nonlinear term in the   NLSE  (\ref{e0}).

The $A_{2}$ integrability conditions  are obtained choosing $j=2$ in the compatibility conditions~(\ref{Lavinia}) with $s=2$ and $s'=3$ as in~(\ref{Turno}).
In this case we have that $f_{2}(2)\in\mathcal P_{4}(1)$ and $f_{3}(2)\in\mathcal P_{5}(1)$ where $\mathcal P_4(1)$ contains 2 dif\/ferent dif\/ferential monomials and $\mathcal P_5(1)$ contains 5 dif\/ferent dif\/ferential monomials, so that $f_{2}(2)$ and $f_{3}(2)$ will be respectively identif\/ied by~2 and~5 complex constants
\begin{gather}
f_{2}(2)\doteq au_{,n_1}(1)|u(1)|^2+b\bar u_{,n_1}(1)u(1)^2,\label{Abruzzo1}\\
f_{3}(2)\doteq\alpha |u(1)|^4u(1)+\beta |u_{,n_1}(1)|^2u(1)+\gamma u_{,n_1}(1)^2\bar u(1)
  +\delta\bar u_{,2n_1}(1)u(1)^2+\varepsilon |u(1)|^2u_{,2n_1}(1).\nonumber
\end{gather}
In this way, eliminating from equation~(\ref{Turno})  the derivatives of $u(1)$ with respect to the slow-times $m_{2}$ and $m_{3}$ using the evolutions (\ref{Valentia1}) with $s=2, 3$ and equating term by term, we obtain that the $A_{2}$ $C$-integrability conditions gives  no constraints on the coef\/f\/icients  $a$ and $b$ appearing in~$f_2(2)$.
 The expression of the coef\/f\/icients $\alpha$, $\beta$, $\gamma$, $\delta$, $\varepsilon$ appearing in $f_3(2)$ in terms of $a$ and $b$ are
\begin{gather}
\alpha= 0,\qquad \beta=-\frac{3i B_{3}b} {B_{2}},\qquad \gamma=-\frac{3i B_{3}a} {2B_{2}},\qquad \delta=0,\qquad \varepsilon=\gamma.\nonumber
\end{gather}
The $A_{3}$ $C$-integrability conditions are derived in a similar way setting $j=3$ in equation~(\ref{Turno}). In this case we have that $f_{2}(3)\in\mathcal P_{5}(2)$ and $f_{3}(3)\in\mathcal P_{6}(2)$ where $\mathcal P_5(2)$ contains 12 dif\/ferent dif\/ferential monomials and $\mathcal P_6(2)$ contains 26 dif\/ferent dif\/ferential monomials, so that $f_{2}(3)$ and~$f_{3}(3)$ will be respectively identif\/ied by 12 and 26 complex constants
\begin{gather}
f_{2}(3)\doteq\tau_{1}|u(1)|^4u(1)+\tau_{2}|u_{,n_1}(1)|^2u(1)+\tau_{3}|u(1)|^2u_{,2n_1}(1)+\tau_{4}\bar u_{,2n_1}(1)u(1)^2 \nonumber\\
\phantom{f_{2}(3)\doteq}{}
+\tau_{5}u_{,n_1}(1)^2\bar u(1)+\tau_{6}u_{,n_1}(2)|u(1)|^2+\tau_{7}\bar u_{,n_1}(2)u(1)^2+\tau_{8}u(2)^2\bar u(1) \label{Lazio4}\\
\phantom{f_{2}(3)\doteq}{}
 +\tau_{9}|u(2)|^2u(1)+\tau_{10}u(2)u_{,n_1}(1)\bar u(1)+\tau_{11}u(2)\bar u_{,n_1}(1)u(1)
  +\tau_{12}\bar u(2)u_{,n_1}(1)u(1),\nonumber\\
f_{3}(3)\doteq\gamma_{1}|u(1)|^4u_{,n_1}(1)+\gamma_{2}|u(1)|^2u(1)^2\bar u_{,n_1}(1)+\gamma_{3}|u(1)|^2u_{,3n_1}(1)
  +\gamma_{4}u(1)^2\bar u_{,3n_1}(1)\nonumber\\
  \phantom{f_{3}(3)\doteq}{}
  +\gamma_{5}|u_{,n_1}(1)|^2u_{,n_1}(1)+\gamma_{6}\bar u_{,2n_1}(1)u_{,n_1}(1)u(1)+
 \gamma_{7}u_{,2n_1}(1)\bar u_{,n_1}(1)u(1)\nonumber\\
 \phantom{f_{3}(3)\doteq}{}
+\gamma_{8}u_{,2n_1}(1)u_{,n_1}(1)\bar u(1)+\gamma_{9}|u(1)|^4u(2)
  +\gamma_{10}|u(1)|^2u(1)^2\bar u(2)+\gamma_{11}\bar u_{,n_1}(1)u(2)^2\nonumber\\
\phantom{f_{3}(3)\doteq}{}
  +\gamma_{12}u_{,n_1}(1)|u(2)|^2
 +\gamma_{13}|u_{,n_1}(1)|^2u(2)+\gamma_{14}|u(2)|^2u(2)+\gamma_{15}u_{,n_1}(1)^{2}\bar u(2) \nonumber\\
 \phantom{f_{3}(3)\doteq}{}
  +\gamma_{16}|u(1)|^2u_{,2n_1}(2)+\gamma_{17}u(1)^2\bar u_{,2n_1}(2)+\gamma_{18}u(2)\bar u_{,2n_1}(1)u(1)
  \nonumber\\
 \phantom{f_{3}(3)\doteq}{}
  +\gamma_{19}u(2)u_{,2n_1}(1)\bar u(1)+\gamma_{20}\bar u(2)u_{,2n_1}(1)u(1)+\gamma_{21}u(2)u_{,n_1}(2)\bar u(1) \nonumber\\
\phantom{f_{3}(3)\doteq}{}
 +\gamma_{22}\bar u(2)u_{,n_1}(2)u(1)+\gamma_{23}u_{,n_1}(2)u_{,n_1}(1)\bar u(1)+\gamma_{24}u_{,n_1}(2)\bar u_{,n_1}(1)u(1) \nonumber\\
 \phantom{f_{3}(3)\doteq}{}
  +\gamma_{25}\bar u_{,n_1}(2)u_{,n_1}(1)u(1)+\gamma_{26}\bar u_{,n_1}(2)u(2)u(1).\nonumber
\end{gather}
Let us eliminate from equation~(\ref{Turno}) with $j=3$ the derivatives of $u(1)$ with respect to the slow-times $m_{2}$ and $m_{3}$ using the evolutions (\ref{Valentia1}) with $s=2, 3$ and the same derivatives of~$u(2)$ using the evolutions (\ref{Valentia2}) with $s=2, 3$. Equating the remaining terms term by term, the $A_{3}$ $C$-integrability conditions turn out to be:
\begin{gather}
\tau_{1}=-\frac{i} {4B_{2}}\left[b\left(\tau_{11}-2\tau_{6}\right)+\bar a\tau_{7}\right],\qquad \bar b\tau_{7}=\frac{1} {2}\left(b-a\right)\left(\tau_{11}+\tau_{10}-\tau_{6}\right)+\bar a\tau_{7},\nonumber\\
a\tau_{8}=b\tau_{8}=0,\qquad a\tau_{9}=b\tau_{9}=0,\qquad \bar a\tau_{12}=a\left(\tau_{10}-\tau_{11}\right)+b\tau_{6}+\bar a\tau_{7},\nonumber\\
\left(\bar b-\bar a\right)\tau_{12}=\left(b-a\right)\tau_{10}.\label{Aurunci}
\end{gather}
Sometimes $a$ and $b$ turn out to be both real. In this case the conditions given in equations ({\ref{Aurunci}}) becomes:
\begin{gather}
R_{1}=\frac{1} {4 B_2}\left[b\left(I_{11}-2I_{6}\right)+aI_{7}\right],\qquad I_{1}=-\frac{1} {4 B_2}\left[b\left(R_{11}-2R_{6}\right)+aR_{7}\right],\nonumber\\
\left(b-a\right)\left(R_{11}+R_{10}-R_{6}-2R_{7}\right)=0,\qquad \left(b-a\right)\left(I_{11}+I_{10}-I_{6}-2I_{7}\right)=0,\nonumber\\
\left(b-a\right)R_{8}=0,\qquad \left(b-a\right)I_{8}=0,\qquad \left(b-a\right)R_{9}=0,\qquad \left(b-a\right)I_{9}=0,\nonumber\\
a\left(R_{12}+R_{11}-R_{10}-R_{7}\right)=bR_{6},\qquad a\left(I_{12}+I_{11}-I_{10}-I_{7}\right)=bI_{6},\nonumber\\
\left(b-a\right)\left(R_{12}-R_{10}\right)=0,\qquad \left(b-a\right)\left(I_{12}-I_{10}\right)=0,\label{Aurunci2}
\end{gather}
where $\tau_i=R_i+i I_i$ for $i=1, \dots, 12$.
The expressions of the $\gamma_{j}$ as functions of the $\tau_{i}$ are:
\begin{gather*}
\gamma_{1}=\frac{3B_{3}} {4B_{2}^2}\left(a\tau_{6}-4i B_{2}\tau_{1}+\bar b\tau_{12}\right),\qquad \gamma_{2}=\frac{3B_{3}} {4B_{2}^2}\left(b\tau_{6}+\bar a\tau_{7}\right),\nonumber\\
\gamma_{3}=-\frac{3i B_{3}\tau_{3}} {2B_{2}},\qquad \gamma_{4}=0,\ \ \ \gamma_{5}=-\frac{3i B_{3}\tau_{2}} {2B_{2}},\qquad \gamma_{6}=-\frac{3i B_{3}\tau_{4}} {B_{2}},\nonumber\\
\gamma_{7}=\gamma_{5},\qquad \gamma_{8}=\gamma_{3}-\frac{3i B_{3}\tau_{5}} {B_{2}},\qquad \gamma_{9}=\gamma_{10}=\gamma_{11}=0,\nonumber\\
\gamma_{12}=-\frac{3i B_{3}\tau_{9}} {2B_{2}},\qquad \gamma_{13}=-\frac{3i B_{3}\tau_{11}} {2B_{2}},\qquad \gamma_{14}=0,\qquad \gamma_{15}=-\frac{3i B_{3}\tau_{12}} {2B_{2}},\nonumber\\
\gamma_{16}=-\frac{3i B_{3}\tau_{6}} {2B_{2}},\qquad \gamma_{17}=\gamma_{18}=0,\qquad \gamma_{19}=-\frac{3i B_{3}\tau_{10}} {2B_{2}},\qquad \gamma_{20}=\gamma_{15},\nonumber\\ \nonumber
\gamma_{21}=-\frac{3i B_{3}\tau_{8}} {B_{2}},\qquad \gamma_{22}=\gamma_{12},\qquad \gamma_{23}=\gamma_{16}+\gamma_{19},\qquad \gamma_{24}=\gamma_{13},\\ \nonumber
\gamma_{25}=-\frac{3i B_{3}\tau_{7}} {B_{2}},\qquad \gamma_{26}=0.
\end{gather*}
The conditions given in equations~(\ref{Aurunci}), (\ref{Aurunci2}) appear to be new. Their importance resides in the fact that a $C$-integrable equation must satisfy those conditions.

\section{Linearizability of the equations of the  Burgers hierarchy}\label{section3}

Taking into account the results of the previous section we can carry out the multiple scale expansion of the equations of the  Burgers hierarchy. To do so we substitute the def\/inition (\ref{Catilina}) into  equations~(\ref{Numa}), (\ref{Pitagora}) and write down the coef\/f\/icients of the various harmonics $\theta$ and of the various orders $j$ of $\epsilon$. When we deal with the dif\/ferential-dif\/ference equation (\ref{Numa}), we have to make the substitutions $\sigma m\rightarrow t$, $\sigma m_i\rightarrow t_i$. This transformation implies that in this case the corresponding coef\/f\/icients $\rho_2$ and $B_2$ will turn out to be $\sigma$-independent. In Appendix we present all relevant equations and here we just present their results.
\begin{proposition}
The differential-difference  equation \eqref{Numa} of the Burgers hierarchy satisfies the~$A_1$  $($and consequently also the $A_2)$ and also the $A_3$ $C$-integrability conditions.
\end{proposition}

\begin{proposition}
The partial difference Burgers-like equation \eqref{Pitagora} reduces for $j=3$ and $\alpha=1$ to a NLSE with a nonlinear complex coefficient $\rho_2$ given by equation \eqref{Roma}. Thus the equation is neither $S$-integrable nor $C$-integrable.
\end{proposition}

\section{Conclusions}\label{section4}

In the present paper we have presented all the steps necessary to apply the perturbative multiple scale expansion to dispersive nonlinear dif\/ferential-dif\/ference or partial dif\/ference equations which may be linearizable. These passages involve the representation of the lattice variables in terms of an inf\/inite set of derivative with respect to the lattice index and the analysis of the higher order of the perturbation which give rise to a set of compatible higher order linear PDE's belonging to the hierarchy of the Schr\"odinger equation. The compatibility of these equations give rise to a linearizability test. We applied the so obtained test to the case of a dif\/ferential-dif\/ference dispersive Burgers equation and its discretization.  It turns out that the Burgers is linearizable (as it should be) but its discretization is neither $S$-integrable nor $C$-integrable. So, ef\/fectively this procedure is able to distinguish between linearizable and non-linearizable equations.

\appendix

  \section{Appendix}

Let us now start performing a multiple scale analysis of the partial dif\/ference  equation (\ref{Pitagora}). We present here the equations we get at the various orders of $\epsilon$ and for the dif\/ferent harmonics~$\theta$.
\begin{itemize}\itemsep=0pt

\item\emph{Order $\epsilon$ and $\theta=0$}: In this case the resulting equation is automatically satisf\/ied.
\item\emph{Order $\epsilon$ and $\theta=1$}: If one requires that $u_{1}^{(1)}\not=0$, one obtains the dispersion relation
\begin{gather}
\sin\left(\omega\sigma\right)=-\frac{\sin\left(\kappa h\right)\sigma} {h}.\label{Armentano}
\end{gather}
\item\emph{Order $\epsilon^2$ and $\theta=0$}: We obtain the evolution
\begin{gather}
\partial_{m_{1}}u_{1}^{(0)}-\frac{\sigma} {h}\partial_{n_{1}}u_{1}^{(0)}=0,\label{Platone}
\end{gather}
which implies that $u_{1}^{(0)}$ depends on the variable $\rho\doteq hn_{1}+\sigma m_{1}$.

\item\emph{Order $\epsilon^2$ and $\theta=1$}: Taking into account the dispersion relation (\ref{Armentano}), we have
\begin{gather}
\partial_{m_{1}}u_{1}^{(1)}-\frac{\sigma} {\cos\left(\omega\sigma\right)}\left[\frac{\cos\left(\kappa h\right)} {h}\partial_{n_{1}}u_{1}^{(1)}-2\sin^2\left(\frac{\kappa h}{2}\right) u_{1}^{(0)}u_{1}^{(1)}\right]=0,\label{GruppodiUr}
\end{gather}
\noindent which implies that $u_{1}^{(1)}$ has the form
\begin{gather}
u_{1}^{(1)}=g\left(\xi,m_{j},j\geq 2\right)\exp\left\{\delta\int_{\rho_{0}}^{\rho}u_{1}^{(0)}\left(\rho^{\prime}\right)d\rho^{\prime}\right\},\qquad \delta\doteq\frac{2\sin^2\left(\kappa h/2\right)} {\left[\cos\left(\kappa h\right)-\cos\left(\omega\sigma\right)\right]},\label{Parise}
\end{gather}
where $\xi\doteq hn_{1}+\frac{\cos\left(\kappa h\right)} {\cos\left(\omega\sigma\right)}\sigma m_{1}$, $g$ is an arbitrary function of its arguments going to zero as $\xi\rightarrow\pm\infty$  and $\rho_{0}$, by a proper redef\/inition of $g$, can always be chosen to be a zero of $u_{1}^{(0)}$  as when $\rho\rightarrow\pm\infty$, $u_{1}^{(0)}\rightarrow 0$ so that there will exists at least one zero.

\item\emph{Order $\epsilon^2$ and $\theta=2$}: Taking into account the dispersion relation (\ref{Armentano}), we have
\begin{gather}
u_{2}^{(2)}=\frac{\left(e^{-i\kappa h}-1\right)h} {2\left[\cos\left(\kappa h\right)-\cos\left(\omega\sigma\right)\right]}\big[u_{1}^{(1)}\big]^2.\label{GiacomoBoni}
\end{gather}
\item\emph{Order $\epsilon^3$ and $\theta=0$}: We have
\begin{gather}
\partial_{m_{1}}u_{2}^{(0)}-\frac{\sigma} {h}\partial_{n_{1}}u_{2}^{(0)}=-\partial_{m_{2}}u_{1}^{(0)}-2\sigma\sin^2\left(\kappa h/2\right)\big[\partial_{n_{1}}+2hu_{1}^{(0)}\big]\big|u_{1}^{(1)}\big|^2.\label{Iuppiter}
\end{gather}
By equation~(\ref{Platone}), the term $\partial_{m_{2}}u_{1}^{(0)}$ is a solution of the left hand side of equation~(\ref{Iuppiter}), hence it is a secular term. As a consequence we have to require that
\begin{gather*}%\label{c}
\partial_{m_{1}}u_{2}^{(0)}-\frac{\sigma} {h}\partial_{n_{1}}u_{2}^{(0)}=-2\sigma\sin^2\left(\kappa h/2\right)\big[\partial_{n_{1}}+2hu_{1}^{(0)}\big]\big|u_{1}^{(1)}\big|^2,\label{c1}\\
\partial_{m_{2}}u_{1}^{(0)}=0.
\end{gather*}
Solving equation~(\ref{c1}) taking into account equation~(\ref{Parise}), we obtain
\begin{gather} \label{Ekatlos}
u_{2}^{(0)}=f\left(\rho,m_{j},j\geq 2\right)-h\delta\cos\left(\omega\sigma\right)\left[|u_{1}^{(1)}|^2 +2\left(1+\delta\right)u_{1}^{(0)}\int_{\xi_{0}}^{\xi}|u_{1}^{(1)}|^2d\xi^{\prime}\right],
\end{gather}
where $f$ is an arbitrary function of its arguments going to zero as $\rho\rightarrow\pm\infty$ and $\xi_{0}$ is an arbitrary value of the variable $\xi$. Let us restrict ourselves for simplicity to the case where there is no dependence at all on $\rho$. If one wants that the harmonic $u_{1}^{(0)}$ depends on $\xi$ and not on $\rho$, from equation~(\ref{Platone}) one has that $\partial_{\xi}u_{1}^{(0)}=0$, so that $u_{1}^{(0)}$ depends on the slow variables $m_{j}$, $j\geq 2$ only. Similarly we have that $\partial_{\xi}f=0$. In this case, in order to satisfy the asymptotic conditions $\lim\limits_{\xi\rightarrow\pm\infty}u_{\gamma}^{(0)}=0$, $\gamma=1$, 2, one has to take $u_{1}^{(0)}=f=0$ (unless we take the fully continuous limit $h\rightarrow 0$, $hn_{1}\doteq x_{1}$, $\sigma\rightarrow 0$, $\sigma m_{1}\doteq t_{1}$ in which $\rho\rightarrow\xi$). Equation~(\ref{Ekatlos}) then becomes
\begin{gather}
u_{2}^{(0)}=-h\delta\cos\left(\omega\sigma\right)\big|u_{1}^{(1)}\big|^2.\label{Caetani}
\end{gather}
\item\emph{Order $\epsilon^3$ and $\theta=1$}: Taking into account the dispersion relation (\ref{Armentano}) and the equations~$u_{1}^{(0)}=0$, (\ref{GruppodiUr}), (\ref{GiacomoBoni}), (\ref{Caetani}), we have
\begin{subequations}
\begin{gather}
\partial_{m_{1}}u_{2}^{(1)}-\frac{\sigma\cos\left(\kappa h\right)} {h\cos\left(\omega\sigma\right)}\partial_{n_{1}}u_{2}^{(1)}=-\partial_{m_{2}}u_{1}^{(1)}-i B_{2}\partial_{\xi}^{2}u_{1}^{(1)}-i\rho_{2}\big|u_{1}^{(1)}\big|^2 u_{1}^{(1)},\label{Amor}\\
B_{2}\doteq\frac{h\sigma\left(h^2-\sigma^2\right)\sin\left(\kappa h\right)} {2\left[\sigma^2\sin^2\left(\kappa h\right)-h^2\right]\cos\left(\omega\sigma\right)}, \label{Orma}\\
\nonumber\rho_{2}\doteq-\frac{2h\sigma\sin^2\left(\omega\sigma/2\right)\left[2\cos\left(\kappa h/2\right)-\cos\left(3\kappa h/2\right)\right]\sin\left(\kappa h/2\right)} {\left[\cos\left(\kappa h\right)-\cos\left(\omega\sigma\right)\right]\cos\left(\omega\sigma\right)}   \\
\phantom{\rho_{2}\doteq}{}
 -\frac{2i h\sigma\sin^2\left(\omega\sigma/2\right)\sin\left(\kappa h/2\right)\sin\left(\kappa h/2\right)} {\left[\cos\left(\kappa h\right)-\cos\left(\omega\sigma\right)\right]\cos\left(\omega\sigma\right)}.\label{Roma}
\end{gather}
\end{subequations}
As a consequence of equation~(\ref{GruppodiUr}) with $u_{1}^{(0)}=0$, the right hand side of equation~(\ref{Amor}) is secular. Hence we have to require that
\begin{subequations}
\begin{gather}
\partial_{m_{1}}u_{2}^{(1)}-\frac{\sigma\cos\left(\kappa h\right)} {h\cos\left(\omega\sigma\right)}\partial_{n_{1}}u_{2}^{(1)}=0,\label{Fauno}\\
i\partial_{m_{2}}u_{1}^{(1)}=B_{2}\partial_{\xi}^{2}u_{1}^{(1)}+\rho_{2}\big|u_{1}^{(1)}\big|^2 u_{1}^{(1)}.\label{Pico}
\end{gather}
\end{subequations}
\end{itemize}
Equation~(\ref{Fauno}) implies that $u_{2}^{(1)}$ also depends on $\xi$  while equation~(\ref{Pico}) is a nonintegrable nonlinear Schr\"odinger equation, as from the definition (\ref{Roma}) we can see that $\rho_{2}$ is a complex coef\/f\/icient. So we can conclude that equation~(\ref{Pitagora}) is not $A_{1}$-integrable.

Let us  perform the multiple scale reduction of the Burgers equation  (\ref{Numa}). Equation (\ref{Numa}) can be always obtained as a semicontinuous limit of equation~(\ref{Pitagora}) def\/ining  the slow times  $t_{j}\doteq\sigma m_{j}$, $j\geq 1$. In such a way we can use in the present calculation the results presented up above.
\begin{itemize}\itemsep=0pt
\item\emph{Order $\epsilon$ and $\theta=0$}: In this case the resulting equation is automatically satisf\/ied.
\item\emph{Order $\epsilon$ and $\theta=1$}: Taking the semi continuous limit of equation~(\ref{Armentano}), one obtains the dispersion relation
\begin{gather}
\omega=-\frac{\sin\left(\kappa h\right)} {h}.\label{Armentano1}
\end{gather}
\item\emph{Order $\epsilon^2$ and $\theta=0$}: Taking the semi continuous limit of equation~(\ref{Platone}), one obtains
\begin{gather}
\partial_{t_{1}}u_{1}^{(0)}-\frac{1} {h}\partial_{n_{1}}u_{1}^{(0)}=0,\label{Platone1}
\end{gather}
 which implies that $u_{1}^{(0)}$ depends on the variable $\rho\doteq hn_{1}+t_{1}$.
\item\emph{Order $\epsilon^2$ and $\alpha=1$}: Taking the semi continuous limit of equation~(\ref{GruppodiUr}), one obtains
\begin{gather}
\partial_{t_{1}}u_{1}^{(1)}-\frac{\cos\left(\kappa h\right)} {h}\partial_{n_{1}}u_{1}^{(1)}+2\sin^2\left(\kappa h/2\right) u_{1}^{(0)}u_{1}^{(1)}=0,\label{GruppodiUr1}
\end{gather}
which implies that $u_{1}^{(1)}$ has the form
\begin{gather}
u_{1}^{(1)}=g^{(1)}\left(\xi,t_{j},j\geq 2\right)\exp\left\{-\int_{\rho_{0}}^{\rho}u_{1}^{(0)}\left(\rho^{\prime}\right)d\rho^{\prime}\right\},\label{Parise1}
\end{gather}
where $\xi\doteq hn_{1}+\cos\left(\kappa h\right) t_{1}$, $g^{(1)}$ is an arbitrary function of its arguments going to zero as $\xi\rightarrow\pm\infty$  and $\rho_{0}$, by a proper redef\/inition of $g$, can always be chosen to be a zero of $u_{1}^{(0)}$.

\item\emph{Order $\epsilon^2$ and $\theta=2$}: Taking the semi continuous limit of equation~(\ref{GiacomoBoni}), one obtains
\begin{gather}
u_{2}^{(2)}=\frac{h} {1-e^{i\kappa h}}u_{1}^{(1)2}.\label{GiacomoBoni1}
\end{gather}

\item\emph{Order $\epsilon^3$ and $\theta=0$}: Taking the semi continuous limit of equation~(\ref{Iuppiter}), one obtains
\begin{eqnarray}
\partial_{t_{1}}u_{2}^{(0)}-\frac{1} {h}\partial_{n_{1}}u_{2}^{(0)}=-\partial_{t_{2}}u_{1}^{(0)}-2\sin^2\left(\kappa h/2\right)\big[\partial_{n_{1}}+2hu_{1}^{(0)}\big]\big|u_{1}^{(1)}\big|^2.\label{Iuppiter1}
\end{eqnarray}
By equation~(\ref{Platone1}), the term $\partial_{t_{2}}u_{1}^{(0)}$ is a solution of the left hand side of equation~(\ref{Iuppiter1}), hence it is a secular term. As a consequence we have to require that
\begin{gather}
\partial_{t_{1}}u_{2}^{(0)}-\frac{1} {h}\partial_{n_{1}}u_{2}^{(0)}=-2\sin^2\left(\kappa h/2\right)\big[\partial_{n_{1}}+2hu_{1}^{(0)}\big]\big|u_{1}^{(1)}\big|^2,\label{c111}\\
\partial_{t_{2}}u_{1}^{(0)}=0.  \nonumber
\end{gather}
Solving equation~(\ref{c111}) taking into account equation~(\ref{Parise1}), we obtain
\begin{gather}
u_{2}^{(0)}=f\left(\rho,t_{j},j\geq 2\right)+h\big|u_{1}^{(1)}\big|^2,\label{Ekatlos1}
\end{gather}
where $f$ is an arbitrary function of its arguments going to zero $\rho \rightarrow
\pm \infty$.

\item\emph{Order $\epsilon^3$ and $\theta=1$}: For simplicity, from now on we require no dependence  on $\rho$\footnote{If $\partial_{\rho}u_{1}^{(0)}\not=0$, $\partial_{\rho}f\not=0$,  we have:
\begin{gather}
\nonumber u_{2}^{(1)}\doteq g^{(2)}\left(n_{1},t_{j},j\geq 1\right)\exp\left\{-\int_{\rho_{0}}^{\rho}u_{1}^{(0)}\left(\rho^{\prime}\right)d\rho^{\prime}\right\},\qquad i\partial_{t_{2}}g^{(1)}=B_{2}\partial_{\xi}^2g^{(1)},  \\
\nonumber g^{(2)}/g^{(1)}=p\left(\xi,t_{j},j\geq 2\right)+h\left[1+\frac{i} {2}\cot\left(\frac{\kappa h} {2}\right)\right]u_{1}^{(0)}+\frac{h} {2}\int_{\rho_{0}}^{\rho}u_{1}^{(0)2}d\rho^{\prime}-\int_{\rho_{0}}^{\rho}f\left(\rho^{\prime}\right)d\rho^{\prime},
\end{gather}
with $p$ arbitrary function of its arguments going to zero as $\xi\rightarrow\pm\infty$.} so that, in order to satisfy the asymptotic conditions, it necessarily follows that
\begin{gather}\label{cc1}
u_{1}^{(0)}=f=0.
\end{gather}
Taking the semi continuous limit of equations~(\ref{Amor}), (\ref{Orma}), one obtains
\begin{gather}
\partial_{t_{1}}u_{2}^{(1)}-\frac{\cos\left(\kappa h\right)} {h}\partial_{n_{1}}u_{2}^{(1)}=-\partial_{t_{2}}u_{1}^{(1)}-i B_{2}\partial_{\xi}^{2}u_{1}^{(1)},\qquad B_{2}\doteq-\frac{h\sin\left(\kappa h\right)} {2}.\label{Amor1}
\end{gather}
As a consequence of equation~(\ref{GruppodiUr1}) (with $u_{1}^{(0)}=0$), the right hand side of equation~(\ref{Amor1}) is secular. Hence we have to require that
\begin{subequations}\label{Dioscuri}
\begin{gather}
\partial_{t_{1}}u_{2}^{(1)}-\frac{\cos\left(\kappa h\right)} {h}\partial_{n_{1}}u_{2}^{(1)}=0,\label{Fauno1}\\
i\partial_{t_{2}}u_{1}^{(1)}-B_{2}\partial_{\xi}^{2}u_{1}^{(1)}=0.\label{Pico1}
\end{gather}
\end{subequations}
Equation~(\ref{Fauno1}) implies that $u_{2}^{(1)}$ depends also on $\xi$  while, contrary to equation~(\ref{Pico}), equation~(\ref{Pico1}) now is a linear Schr\"odinger equation, ref\/lecting the $C$-integrability of equation~(\ref{Numa}).

\item\emph{Order $\epsilon^3$ and $\alpha=2$}: Taking into account the dispersion relation (\ref{Armentano1}), the fact that $u_{1}^{(0)}=0$ and the equations~(\ref{GruppodiUr1}), (\ref{GiacomoBoni1}), we have
\begin{gather}
 u_{3}^{(2)}=h\left[\frac{2} {1-e^{i\kappa h}}u_{2}^{(1)}-\frac{h} {4\sin^2\left(\kappa h/2\right)}\partial_{\xi}u_{1}^{(1)}\right]u_{1}^{(1)}.\label{Luce}
\end{gather}

\item\emph{Order $\epsilon^3$ and $\theta\!=3$}: Taking into account the dispersion relation (\ref{Armentano1}) and equa\-tion~(\ref{GiacomoBoni1}), we obtain
\begin{gather}
 u_{3}^{(3)}=\left(\frac{h} {1-e^{i\kappa h}}\right)^2\big[u_{1}^{(1)}\big]^3.\label{Saturnia}
\end{gather}

\item\emph{Order $\epsilon^4$ and $\theta=0$}: Taking into account equations~(\ref{GiacomoBoni1}), (\ref{Ekatlos1}), (\ref{cc1}), (\ref{Pico1}), we get
\begin{gather}\label{PietrodeAngelis}
\partial_{t_{1}}u_{3}^{(0)}-\frac{1} {h}\partial_{n_{1}}u_{3}^{(0)}=h\partial_{\xi}\left[\frac{h} {2}\partial_{\xi}\big|u_{1}^{(1)}\big|^2-2\sin^2\left(\kappa h/2\right)\big(u_{2}^{(1)}u_{1}^{(-1)}+u_{2}^{(-1)}u_{1}^{(1)}\big)\right].
\end{gather}
Solving equation~(\ref{PietrodeAngelis}), we obtain
\begin{gather}\label{Pentalfa}
u_{3}^{(0)}=\tau\left(\rho,t_{j},j\geq 2\right)+h\big(u_{2}^{(1)}u_{1}^{(-1)}+u_{2}^{(-1)}u_{1}^{(1)}\big)-\frac{h^2} {4\sin^2\left(\kappa h/2\right)}\partial_{\xi}\big|u_{1}^{(1)}\big|^2,
\end{gather}
where $\tau$ is an arbitrary function of its arguments going to zero as $\rho\rightarrow\pm\infty$. As usually, if we don't want any dependence at all from $\rho$ but only on $\xi$, in order to satisfy the asymptotic conditions $\lim\limits_{\xi\rightarrow\pm\infty}u_{3}^{(0)}=0$, we have to take
\begin{gather}\label{cc2}
\tau=0
\end{gather}
 (unless we take the fully continuous limit).
\item\emph{Order $\epsilon^4$ and $\theta=1$}: Taking into account the dispersion relation (\ref{Armentano1}) and equa\-tions (\ref{GiacomoBoni1}), (\ref{Ekatlos1}), (\ref{cc1}), (\ref{Luce}), (\ref{Pentalfa}), (\ref{cc2}), we get
\begin{gather}
\nonumber\partial_{t_{1}}u_{3}^{(1)}-\frac{\cos\left(\kappa h\right)} {h}\partial_{n_{1}}u_{3}^{(1)}=-\partial_{t_{3}}u_{1}^{(1)}
-\partial_{t_{2}}u_{2}^{(1)}-B_{3}\partial_{\xi}^3u_{1}^{(1)}-i B_{2}\partial_{\xi}^2u_{2}^{(1)},\\
B_{3}\doteq-\frac{h^2\cos\left(\kappa h\right)} {6}.\label{Talao}
\end{gather}
As a consequence of the equations~(\ref{GruppodiUr1}), (\ref{Fauno1}) and as $u_{1}^{(0)}=0$, the right hand side of equation~(\ref{Talao}) is secular, so that
\begin{subequations} \label{Dioscuri1}
\begin{gather}
\partial_{t_{1}}u_{3}^{(1)}-\frac{\cos\left(\kappa h\right)} {h}\partial_{n_{1}}u_{3}^{(1)}=0,\label{Mongenet}  \\
\partial_{t_{2}}u_{2}^{(1)}+i B_{2}\partial_{\xi}^2u_{2}^{(1)}=-\partial_{t_{3}}u_{1}^{(1)}-B_{3}\partial_{\xi}^3u_{1}^{(1)}.\label{Ribulsi}
\end{gather}
\end{subequations}
The f\/irst relation implies that $u_{3}^{(1)}$ also depends on $\xi$  while  the second one, as a consequence of equation~(\ref{Pico1}), implies that the right hand side is secular, so that
\begin{subequations}
\begin{gather}
i\partial_{t_{2}}u_{2}^{(1)}-B_{2}\partial_{\xi}^2u_{2}^{(1)}=0,\label{Acca}\\
\partial_{t_{3}}u_{1}^{(1)}+B_{3}\partial_{\xi}^3u_{1}^{(1)}=0.\label{Geronta}\ \
\end{gather}
\end{subequations}
 Equation~(\ref{Acca}), as one can see from the def\/inition (\ref{Abruzzo1}), has a forcing term $f_{2}(2)$ with coef\/f\/icients $a=b=0$.

\item\emph{Order $\epsilon^4$ and $\theta=2$}: Taking into account the dispersion relation (\ref{Armentano1}) and the equations~(\ref{GruppodiUr1}), (\ref{GiacomoBoni1}), (\ref{Ekatlos1}), (\ref{cc1}), (\ref{Dioscuri}), (\ref{Luce}), (\ref{Saturnia}), we get
\begin{gather}
u_{4}^{(2)}=-h^3\left[\frac{1} {\left(1-e^{i\kappa h}\right)^2}u_{1}^{(1)}|u_{1}^{(1)}|^2+\frac{i\cos\left(\kappa h/2\right)} {8\sin^3\left(\kappa h/2\right)}\partial_{\xi}^2u_{1}^{(1)}\right]u_{1}^{(1)} \nonumber\\
\phantom{u_{4}^{(2)}=}{}
-\frac{h^2} {4\sin^2\left(\kappa h/2\right)}\partial_{\xi}\big(u_{1}^{(1)}u_{2}^{(1)}\big)+\frac{h} {1-e^{i\kappa h}}\big(u_{2}^{(1)2}+2u_{1}^{(1)}u_{3}^{(1)}\big).\label{Atanor}
\end{gather}
\item\emph{Order $\epsilon^4$ and $\theta=3$}: Taking into account the dispersion relation (\ref{Armentano1}) and the equations~(\ref{GruppodiUr1}), (\ref{GiacomoBoni1}), (\ref{cc1}), (\ref{Luce}), (\ref{Saturnia}), we get
\begin{gather*}
u_{4}^{(3)}=\left(\frac{h} {1-e^{i\kappa h}}u_{1}^{(1)}\right)^2\left(3u_{2}^{(1)}+\frac{2he^{i\kappa h}} {1-e^{i\kappa h}}\partial_{\xi}u_{1}^{(1)}\right).
\end{gather*}
\item\emph{Order $\epsilon^4$ and $\theta=4$}: Taking into account the dispersion relation (\ref{Armentano1}) and the equations~(\ref{GiacomoBoni1}), (\ref{Saturnia}), we get
\[
u_{4}^{(4)}=\left(\frac{h} {1-e^{i\kappa h}}\right)^3\big[u_{1}^{(1)}\big]^4.
\]
\item\emph{Order $\epsilon^5$ and $\theta=0$}: Taking into account equations~(\ref{GiacomoBoni1}), (\ref{Ekatlos1}), (\ref{cc1}), (\ref{Pico1}), (\ref{Luce}), (\ref{cc2}), (\ref{Pentalfa}), (\ref{Dioscuri1}), we get
\begin{gather}
\partial_{t_{1}}u_{4}^{(0)}-\frac{1} {h}\partial_{n_{1}}u_{4}^{(0)}=h\partial_{\xi}\Big\{-2\sin^2\left(\kappa h/2\right)\big(u_{1}^{(1)}u_{3}^{(-1)}+u_{1}^{(-1)}u_{3}^{(1)}+|u_{2}^{(1)}|^2\big) \nonumber\\
\nonumber\left.
\qquad{}+\left[4\sin^2\left(\kappa h/2\right)-1\right]\frac{h^2} {2}\big|u_{1}^{(1)}\big|^4+\frac{h} {2}\partial_{\xi}\big(u_{1}^{(1)}u_{2}^{(-1)}+u_{1}^{(-1)}u_{2}^{(1)}\big)\right.\\
\qquad{}
+\frac{i h^2} {4}\cot\left(\kappa h/2\right)\partial_{\xi}\big(u_{1}^{(-1)}\partial_{\xi}u_{1}^{(1)}-u_{1}^{(1)}\partial_{\xi}u_{1}^{(-1)}\big)\Big\}.
\label{Rumon}
\end{gather}
Solving equation~(\ref{Rumon}), we obtain
\begin{gather}
u_{4}^{(0)}=\Theta\left(\rho,t_{j},j\geq 2\right)+h\big(u_{1}^{(1)}u_{3}^{(-1)}+u_{1}^{(-1)}u_{3}^{(1)}+|u_{2}^{(1)}|^2\big) \nonumber\\
\nonumber
\phantom{u_{4}^{(0)}=}{}
+\frac{h^2} {4\sin^2\left(\kappa h/2\right)}\left\{\partial_{\xi}\big(u_{1}^{(1)}u_{2}^{(-1)}+u_{1}^{(-1)}u_{2}^{(1)}\big)+\left[4\sin^2\left(\kappa h/2\right)-1\right]h\big|u_{1}^{(1)}\big|^4\right\} \\
\phantom{u_{4}^{(0)}=}{}
-\frac{i h^3\cos\left(\kappa h/2\right)} {8\sin^3\left(\kappa h/2\right)}\partial_{\xi}\big(u_{1}^{(-1)}\partial_{\xi}u_{1}^{(1)}-u_{1}^{(1)}\partial_{\xi}u_{1}^{(-1)}\big),
\label{Ignis}
\end{gather}
where $\Theta$ is an arbitrary function of its arguments going to zero as $\rho\rightarrow\pm\infty$. As usually, if we don't want any dependence from $\rho$ but only on $\xi$, in order to satisfy the asymptotic conditions $\lim\limits_{\xi\rightarrow\pm\infty}u_{4}^{(0)}=0$, we have to take
\[
\Theta=0
\]
(unless we take the fully continuous limit).

\item\emph{Order $\epsilon^5$ and $\theta=1$}: Taking into account the dispersion relation (\ref{Armentano1}) and equations $u_{1}^{(0)}=f=\tau=\Theta=0$, together with the equations (\ref{GiacomoBoni1}), (\ref{Ekatlos1}), (\ref{Luce}), (\ref{Saturnia}), (\ref{Pentalfa}), (\ref{Atanor}), (\ref{Ignis}), we get
\begin{gather}
\partial_{t_{1}}u_{4}^{(1)}-\frac{\cos\left(\kappa h\right)} {h}\partial_{n_{1}}u_{4}^{(1)}=-\partial_{t_{4}}u_{1}^{(1)}-\partial_{t_{3}}u_{2}^{(1)}-\partial_{t_{2}}u_{3}^{(1)}-i B_{2}\partial_{\xi}^2u_{3}^{(1)}-B_{3}\partial_{\xi}^3u_{2}^{(1)} \nonumber\\
\qquad{} +i B_{4}\partial_{\xi}^4u_{1}^{(1)}+\zeta\left[u_{1}^{(1)}\big|\partial_{n_{1}}u_{1}^{(1)}\big|^2+u_{1}^{(-1)}
\big(\partial_{n_{1}}u_{1}^{(1)}\big)^2+u_{1}^{(1)2}\partial_{n_{1}}^2u_{1}^{(-1)}\right],
\label{ScholaItalica}
\\
\nonumber B_{4}\doteq\frac{h^3\sin\left(\kappa h\right)} {24},\qquad \zeta\doteq\frac{h\left[1+\cos\left(\kappa h\right)+3i\sin\left(\kappa h\right)\right]} {e^{i\kappa h}-1}.
\end{gather}
As a consequence of the equations~(\ref{GruppodiUr1}), (\ref{Fauno1}), (\ref{Mongenet})  and of $u_{1}^{(0)}=0$, the right hand side of equation~(\ref{ScholaItalica}) is secular, so that
\begin{gather}
\nonumber\partial_{t_{1}}u_{4}^{(1)}-\frac{\cos\left(\kappa h\right)} {h}\partial_{n_{1}}u_{4}^{(1)}=0, \\
\partial_{t_{2}}u_{3}^{(1)}+i  B_{2}\partial_{\xi}^2u_{3}^{(1)}=-\partial_{t_{4}}u_{1}^{(1)}-\partial_{t_{3}}u_{2}^{(1)}
-B_{3}\partial_{\xi}^3u_{2}^{(1)}\nonumber \\
\qquad{}+ i B_{4}\partial_{\xi}^4u_{1}^{(1)}+\zeta\left[u_{1}^{(1)}
\big|\partial_{n_{1}}u_{1}^{(1)}\big|^2+u_{1}^{(-1)}\big(\partial_{n_{1}}u_{1}^{(1)}\big)^2
+u_{1}^{(1)2}\partial_{n_{1}}^2u_{1}^{(-1)}\right].\label{Lebano}
\end{gather}
The f\/irst relation tells us that $u_{4}^{(1)}$ depends on $\xi$ too while in the second one, as a consequence of equations~(\ref{Pico1}), (\ref{Acca}), the f\/irst our terms in the right hand side of equation~(\ref{Lebano}) are secular, so that
\begin{subequations}
\begin{gather}\label{GentiluomoGabino}
\partial_{t_{2}}u_{3}^{(1)}+i B_{2}\partial_{\xi}^2u_{3}^{(1)}=\zeta\left[u_{1}^{(1)}\big|\partial_{n_{1}}u_{1}^{(1)}\big|^2+u_{1}^{(-1)}
\big(\partial_{n_{1}}u_{1}^{(1)}\big)^2+u_{1}^{(1)2}\partial_{n_{1}}^2u_{1}^{(-1)}\right], \\
\partial_{t_{3}}u_{2}^{(1)}+B_{3}\partial_{\xi}^3u_{2}^{(1)}=-\partial_{t_{4}}u_{1}^{(1)}+i B_{4}\partial_{\xi}^4u_{1}^{(1)}.\label{Sebezio}
\end{gather}
\end{subequations}
\noindent As a consequence of equation~(\ref{Geronta}), the right hand side of equation~(\ref{Sebezio}) is secular so that
\begin{gather*}
 \partial_{t_{3}}u_{2}^{(1)}+B_{3}\partial_{\xi}^3u_{2}^{(1)}=0,\qquad
 \partial_{t_{4}}u_{1}^{(1)}-i B_{4}\partial_{\xi}^4u_{1}^{(1)}=0.
\end{gather*}
 Equation~(\ref{GentiluomoGabino}), as one can see from the def\/inition (\ref{Lazio4}) and taking into account that $a=b=0$, has a forcing term $f_{2}(3)$ that respects all the $A_{3}$ $C$-integrability conditions~(\ref{Aurunci}).

\end{itemize}

\subsection*{Acknowledgements}

The authors have been partly supported by the Italian Ministry of Education and Research, PRIN ``Nonlinear waves: integrable f\/inite dimensional reductions and discretizations'' from 2007 to 2009 and PRIN ``Continuous and discrete nonlinear integrable evolutions: from water waves to symplectic maps'' from 2010.

\pdfbookmark[1]{References}{ref}
\LastPageEnding


\begin{thebibliography}{99}

\footnotesize\itemsep=0pt

\bibitem{Ag}
Agrotis M., Lafortune S., Kevrekidis P.G.,
On a discrete version of the Korteweg--de Vries equation,
{\it Discrete Contin. Dyn. Syst.}   (2005), suppl., 22--29.

\bibitem{burgers}
Burgers J.M.,
A mathematical model illustrating the theory of turbulence,
\href{http://dx.doi.org/10.1016/S0065-2156(08)70100-5}{{\it Adv. Appl. Mech.}} {\bf 1} (1948), 171--199.

\bibitem{ca}
Calogero F.,
Why are certain nonlinear PDEs both widely applicable and integrable?,
in What is Integrability?, Editor   V.E.~Zakharov, {\it Springer Ser. Nonlinear Dynam.},  Springer, Berlin, 1991, 1--62.

\bibitem{ce}
 Calogero F., Eckhaus W.,
 Necessary conditions for integrability of nonlinear PDEs,
\href{http://stacks.iop.org/0266-5611/3/L27}{{\it Inverse Problems}}  {\bf 3}  (1987), L27--L32.\\
Calogero F., Eckhaus W.,
Nonlinear evolution equations, rescalings, model PDEs and their integrability.~I,
\href{http://stacks.iop.org/0266-5611/3/229}{{\it Inverse Problems}}  {\bf 3}  (1987), 229--262.\\
 Calogero F., Eckhaus  W.,
  Nonlinear evolution equations, rescalings, model PDEs and their integrability.~II,
\href{http://stacks.iop.org/0266-5611/4/11}{{\it Inverse Problems}} {\bf  4} (1987), 11--33.\\
 Calogero F., Degasperis A., Ji X-D.,
 Nonlinear Schr\"odinger-type equations from multiple scale reduction of PDEs. I.~Systematic derivation,
\href{http://dx.doi.org/10.1063/1.1287644}{{\it J. Math. Phys.}} {\bf 41} (2000), 6399--6443.\\
Calogero F., Degasperis A., Ji X-D.,
Nonlinear Schr\"odinger-type equations from multiple scale reduction of PDEs.
II.~Necessary conditions of integrability for real PDEs,
\href{http://www.ams.org/leavingmsn?url=http://dx.doi.org/10.1063/1.1366296}{{\it J. Math. Phys.}} {\bf 42} (2001),  2635--2652.\\
 Calogero F., Maccari A.,
 Equations of nonlinear Schr\"odinger type in $1+1$ and $2+1$ dimensions obtained from integrable PDEs,
 in  Inverse Problems: an Interdisciplinary Study (Montpellier, 1986),
 {\it Adv. Electron. Electron Phys.}, Suppl.~19, Editors C.P.~Sabatier, Academic Press, London, 1987, 463--480.

\bibitem{cole}
 Cole J.D.,
 On a quasi-linear parabolic equation occurring in aerodynamics,
 {\it Quart. Appl. Math.} {\bf 9} (1951), 225--236.

\bibitem{degasperisprivate}
Degasperis A., Private communication.

\bibitem{dp1}
Degasperis A.,  Holm D.D.,   Hone A.N.I.,
 A new integrable equation with peakon solutions,
  {\it Teoret. Mat. Fiz.} {\bf 133} (2002), 170--183 (English transl.: \href{http://dx.doi.org/10.1023/A:1021186408422}{{\it Theoret. and Math. Phys.}} {\bf 133} (2002), 1463--1474).

\bibitem{DMS}
 Degasperis A., Manakov S.V., Santini P.M.,
 Multiple-scale perturbation beyond the nonlinear Schr\"odinger equation.~I,
\href{http://dx.doi.org/10.1016/S0167-2789(96)00179-0}{{\it Phys. D}} {\bf 100} (1997), 187--211.

\bibitem{DP}
 Degasperis A., Procesi M.,
 Asymptotic integrability,
in  Symmetry and Perturbation Theory, SPT98 (Rome, 1998),  Editors A.~Degasperis and G.~Gaeta,
World Sci. Publ., River Edge, NJ, 1999, 23--37. \\
 Degasperis A.,
 Multiscale expansion and integrability of dispersive wave equations,
 in 	 Integrability, Editor A.V.~Mikhailov, Springer, Berlin, 2009,  215--244.

\bibitem{HLPS}
 Hernandez Heredero R., Levi D., Petrera M., Scimiterna C.,
 Multiscale expansion of the lattice potential KdV equation on functions of an inf\/inite slow-varyness order,
\href{http://dx.doi.org/10.1088/1751-8113/40/34/F02}{{\it J.~Phys.~A: Math. Theor.}} {\bf 40} (2007), F831--F840,
\href{http://www.arxiv.org/abs/0706.1046}{arXiv:0706.1046}.

\bibitem{HLPS2}
 Hernandez Heredero R., Levi D., Petrera M., Scimiterna C.,
 Multiscale expansion on the lattice and integrability of partial dif\/ference equations,
\href{http://dx.doi.org/10.1088/1751-8113/41/31/315208}{{\it J.~Phys.~A: Math. Theor.}} {\bf 41} (2008),  315208, 12~pages,
\href{http://www.arxiv.org/abs/0710.5299}{arXiv:0710.5299}.

\bibitem{HLPS3}
 Hernandez Heredero R., Levi D., Petrera M., Scimiterna C.,
 Multiscale expansion and integrability pro\-per\-ties of the lattice potential KdV equation,
\href{http://dx.doi.org/10.2991/jnmp.2008.15.s3.31}{{\it J.~Nonlinear Math. Phys.}} {\bf 15} (2008), suppl.~3, 323--333,
\href{http://www.arxiv.org/abs/0709.3704}{arXiv:0709.3704}.

\bibitem{viallet}
Hietarinta J., Viallet C.,
Singularity conf\/inement and chaos in discrete systems,
\href{http://dx.doi.org/10.1103/PhysRevLett.81.325}{{\it Phys. Rev. Lett.}} {\bf 81} (1998), 325--328,
\href{http://www.arxiv.org/abs/solv-int/9711014}{solv-int/9711014}.

\bibitem{hopf}
Hopf E.,
The partial dif\/ferential equation $u_t + uu_x = u_{xx}$,
\href{http://dx.doi.org/10.1002/cpa.3160030302}{{\it Comm. Pure Appl. Math.}} {\bf  3} (1950), 201--230.

\bibitem{MK}
 Kodama Y., Mikhailov A.V.,
 Obstacles to asymptotic integrability,
 in  Algebraic Aspects of Integrable Systems,
 {\it Progr. Nonlinear Differential Equations Appl.}, Vol.~26,
 Birkh\"auser Boston, Boston, MA, 1997, 173--204.\\
 Hiraoka Y., Kodama Y.,
 Normal form and solitons,  in 	 Integrability, Editor A.V.~Mikhailov, Springer, Berlin, 2009,  175--214,
\href{http://www.arxiv.org/abs/nlin.SI/0206021}{nlin.SI/0206021}.

\bibitem{lm}
Leon J., Manna M.,
Multiscale analysis of discrete nonlinear evolution equations,
\href{http://dx.doi.org/10.1088/0305-4470/32/15/012}{{\it J.~Phys.~A: Math. Gen.}}  {\bf 32} (1999), 2845--2869,
\href{http://www.arxiv.org/abs/solv-int/9902005}{solv-int/9902005}.

\bibitem{levi}
 Levi D.,
 Multiple-scale analysis of discrete nonlinear partial dif\/ference equations: the reduction of the lattice potential KdV,
\href{http://dx.doi.org/10.1088/0305-4470/38/35/005}{{\it J.~Phys.~A: Math. Gen.}} {\bf 38} (2005), 7677--7689,
\href{http://www.arxiv.org/abs/nlin.SI/0505061}{nlin.SI/0505061}.

\bibitem{LevHer}
 Levi D., Hernandez Heredero R.,
 Multiscale analysis of discrete nonlinear evolution equations: the reduction of the dNLS,
\href{http://dx.doi.org/10.2991/jnmp.2005.12.s1.36}{{\it J. Nonlinear Math. Phys.}} {\bf 12} (2005), suppl.~1, 440--448.

\bibitem{lp}
 Levi D.,  Petrera M.,
 Discrete reductive perturbation technique,
\href{http://dx.doi.org/10.1063/1.2190776}{{\it J.~Math. Phys.}} {\bf  47} (2006), 043509, 20~pages,
\href{http://www.arxiv.org/abs/math-ph/0510084}{math-ph/0510084}.

\bibitem{31}
 Levi D., Petrera M.,
 Continuous symmetries of the lattice potential KdV equation,
\href{http://dx.doi.org/10.1088/1751-8113/40/15/006}{{\it J. Phys.~A: Math. Theor.}}  {\bf 40} (2007), 4141--4159,
\href{http://www.arxiv.org/abs/math-ph/0701079}{math-ph/0701079}.

\bibitem{BLR}
Levi D., Ragnisco O., Bruschi M.,
Continuous and discrete matrix Burgers' hierarchies,
\href{http://dx.doi.org/10.1007/BF02721683}{{\it Nuovo Cimento~B~(11)}} {\bf  74} (1983), 33--51.

\bibitem{ls2009}
Levi D., Scimiterna C.,
The Kundu--Eckhaus equation and its discretizations,
\href{http://dx.doi.org/10.1088/1751-8113/42/46/465203}{{\it J.~Phys.~A: Math. Theor.}} {\bf 42}  (2009), 465203, 8~pages,
\href{http://www.arxiv.org/abs/0904.4844}{arXiv:0904.4844}.

\bibitem{ramani}
Ramani A., Grammaticos B., Tamizhmani K.M.,
Painlev\'e analysis and singularity conf\/inement: the ultimate conjecture,
\href{http://dx.doi.org/10.1088/0305-4470/26/2/005}{{\it J. Phys. A: Math. Gen.}} {\bf 26} (1993), L53--L58.

\bibitem{santini2010}
Santini P.M.,
The multiscale expansions of dif\/ference equations in the small lattice spacing regime, and a~vicinity and integrability test.~I,
\href{http://dx.doi.org/10.1088/1751-8113/43/4/045209}{{\it J.~Phys.~A: Math. Theor.}} {\bf 43} (2010),  045209, 27~pages,
\href{http://www.arxiv.org/abs/0908.1492}{arXiv:0908.1492}.

\bibitem{tS} Scimiterna C.,
Multiscale techniques for nonlinear dif\/ference equations, Ph.D. Thesis,  Roma Tre University, 2009.

\bibitem{Schoo}
 Schoombie S.W.,
 A discrete multiscales analysis of a discrete version of the Korteweg--de Vries equation,
\href{http://dx.doi.org/10.1016/0021-9991(92)90042-W}{{\it J.~Comp. Phys.}} {\bf 101} (1992), 55--70.

\bibitem{wh} Whitham G.B.,
 Linear and nonlinear waves, Wiley-Interscience, New York, 1974.


\end{thebibliography}
\end{document}